\documentclass[12pt,preprint]{aastex}
\usepackage{amsmath}                
\usepackage{amsfonts}               
\usepackage{amssymb}                
\def \s{~\rm{s}}
\def \km{~\rm{km}}

\def \K{~\rm{K}}

\def \AU{~\rm{AU}}

\def \yr{~\rm{yr}}

\begin{document}

\title{THE SOURCE OF THE HELIUM VISIBLE LINES IN ETA CARINAE}

\author{Amit Kashi\altaffilmark{1} and Noam Soker\altaffilmark{1}}

\altaffiltext{1}{Department of Physics, Technion$-$Israel
Institute of Technology, Haifa 32000 Israel;
kashia@physics.technion.ac.il;
soker@physics.technion.ac.il.}

\begin{abstract}
We assume that the helium-I lines emitted by the massive binary system $\eta$ Carinae are
formed in the acceleration zone of the less-massive secondary star.
We calculate the Doppler shift of the lines as a function of orbital phase and of
several parameters of the binary system.
We find that a good fit is obtained if the helium lines are formed in the
region where the secondary wind speed is $v_{\rm zone}= 430 \km \s^{-1}$.
The acceptable binary eccentricity is in the range $0.90 \la e \la 0.95$,
and the inclination angle (the angle between a line perpendicular to the orbital
plane and the line of sight) is in the range $40^\circ \la i \la 55 ^\circ$.
Lower values of $e$ require higher values of $i$, and vice versa.
The binary system is oriented such that the secondary star is in our
direction (closer to us) during periastron passage.
The orbital motion can account in part to the Doppler shift of the
peak in X-ray emission.
\end{abstract}

\section{INTRODUCTION}
\label{sec:intro}

If it was not for its binary companion (the secondary $\eta~$Car~B star),
the primary massive star ($\eta$~Car~A) would be a `boring' star.
It is the binary companion that most likely shaped the Homunculus (Soker 2003; 2007b)$-$the
bipolar nebula around $\eta$ Car that was ejected during the
twenty years long Great Eruption (Davidson \& Humphreys 1997).
The orbital motion is also behind the $5.54 \yr$ year periodic variation in the light
curve of $\eta$ Car, from the radio (Duncan \& White 2003) to the X-ray band (Corcoran 2005),
and in all wavelengths in between (e.g., Damineli 1996; Damineli et al.\ 1998, 2000;
Whitelock et al. 2004; Smith et al. 2004; Abraham et al. 2005;
Hillier et al. 2006; Nielsen et al. 2007).

For some emission bands, and emission and absorption lines, there is a consensus
as to their origin, for other there is a dispute.
It is agreed that the hard X-ray emission results from the hot postshock secondary stellar wind
(Corcoran et al. 2001; Pittard \& Corcoran 2002; Akashi et al. 2006; Hamaguchi et al. 2007),
that most of the radio emission comes from extended ($\sim 10^4 AU$)  circumbinary
ionized gas (Cox et al. 1995; Duncan \& White 2003; Kashi \& Soker 2007),
and the source of the hard UV radiation is $\eta~$Car~B (Verner et al. 2005;
Duncan \& White 2003; Soker 2007a).
The location of the plasma emitting the He~II~$\lambda 4686$\AA\ line is in dispute.
Steiner \& Damineli (2004) and Martin et al.\ (2006) assumed that the He~II emission
originates from photoionized regions near the X-ray shock fronts of the two winds.
Soker \& Behar (2006), on the other hand, postulate that this line originates in
the acceleration zone of the secondary wind.

In the present paper we dispute the recent suggestion made by Nielsen et al. (2007)
that the HeI lines originate in the primary wind.
This is described in section 2.
In section 3 we describe the blueshift resulting from the orbital motion.
In section 4 we analyze the velocity of the lines and try to constrain the
orientation angle of the semimajor axis and the inclination of the orbital plane.
The orientation angle of the semi-major axis of $\eta$ Car is another point of dispute.
We summarize in section 5.

\section{THE HELIUM-I VISIBLE LINES}
\label{helium}

The He~I lines ($\lambda$5877, 6680, 7067, 7283) are formed by transitions between
excited levels in neutral helium.
The main properties of these lines in $\eta$ Car are as follows (Nielsen et al. 2007).
The lines are observed both in absorption and emission, and have P-Cygni type profiles
at many orbital phases of $\eta$ car.
The velocity measured at the minimum of the line profile (maximum absorption) relative
to the system velocity along the line of sight is blueshifted at all phases.
After periastron passage the velocity is $\sim -300 \km \s^{-1}$.
The magnitude of the velocity then increases, reaching $\sim -500 \km \s^{-1}$ at phase
$\sim -0.05$ ($\sim 100$ days before periastron).
Very close to periastron passage there is a sharp rise in the magnitude of the velocity,
probably due to a second wind component (Nielsen et al. 2007), to $\sim -600 \km \s^{-1}$.
During the spectroscopic event the absorption line velocity rapidly shifts from
$\sim -600 \km \s^{-1}$ to $\sim -300 \km \s^{-1}$.

The equivalent widths of the lines in emission are more or less constant along
most of the orbit (beside some fluctuations).
At phase $\sim -0.1$ they start to drop, reaching a minimum at phase $\sim 0$,
where they stay for the spectroscopic event of $\sim 10~$weeks.
The equivalent widths in absorption are low during most of the orbit.
At phase $\sim -0.1$ they start to rise reaching a maximum just before
phase zero. During the spectroscopic event they sharply drop, partially recovering
after the event (for more detail see Nielsen et al. 2007).
The general behavior of the HeI lines in absorption as reported by Nielsen et al. (2007)
is qualitatively similar, in both velocity and equivalent width,
to the behavior of the He~II $\lambda$4686 emission line
as reported by Steiner \& Damineli (2004; see also Martin et al. 2006).

Nielsen et al. (2007) proposed that the HeI emission and absorption originate in
the primary wind, where the helium is ionized (or excited) by the secondary stellar
radiation.
Luminous blue variable stars similar in properties to $\eta$ Car A
are known to have HeI lines with P-Cygni type profiles, e.g., P-Cygni itself
(Stahl et al. 1993; Najarro et al. 1997), without any external ionizing source.
Therefore, it is possible in principle that the source of the HeI lines is
indeed the primary wind.
However, there are some difficulties with this possibility:
(1) What cause the monotonic variation in the blueshift along the orbit?
(2) Why do the equivalent widths of absorption and emission drop for several weeks
near periastron passage?
{{{ (3) The most severe problem is the behavior of the He I emission.
Not only the absorption, put the entire emission part
changes its Doppler shift with phase. The width of the emission peak does not change much.
This is expected if the line is formed by the orbiting secondary star, as we propose.
To the contrary, if the line originates in the primary wind, then as regions further out
become the source of the line, with their higher velocity, so is the emission peak suppose
to be wider. This is not observed (fig. 2 of Nielsen et al. 2007). }}}
Considering these difficulties, we follow Soker \& Behar (2006) who attributed the
He~II $\lambda$4686 line to the wind from the $\eta$ Car B, and propose the same for
the HeI lines. Namely, we argue that the He I lines originate in the acceleration
zone of the secondary wind.

WR and O stars are known to have HeII and HeI lines (e.g., Crowther \& Bohannan  1997).
The three stars studied by Crowther \& Bohannan  (1997) are somewhat cooler than $\eta$ Car B,
but the main factor is the mass loss rate (Crowther \& Bohannan  1997), which is similar
to that of $\eta$ Car B.
These three stars all show the He~II $\lambda$4686 line in emission, and
the He~I $\lambda$5876 line in emission and absorption with equivalent widths of
$\sim 0.25-4$~\AA.
These stars have He~I $\lambda$6678 emission, and one has this
line in absorption as well (Crowther \& Bohannan  1997).
The minimum (maximum absorption) in the P-Cygni line profiles of the HeI lines
of O stars occurs typically at a velocity of $\sim -0.1$ to $-0.3$ times the wind's
terminal speed (Prinja et al. 1996, 2001, 2006).
For the $\eta$ Car B this range is $\sim 300-900 \km \s^{-1}$,
in accord with the observations of Nielsen et al. (2007).
We conclude that the HeI line profiles of O stars are similar enough to the line profiles
observed from $\eta$ Car that we may attribute these lines to the wind from
the secondary star.

The fast temporal variability in the intensity and speed of these lines from
O-stars (e.g. Prinja et al. 2006) is evidence of their sensitivity to the
wind properties. Therefore, small changes in the wind properties might result
in large changes in the observed lines profiles.
Soker \& Behar (2006) attributed the large changes in the properties of the
He~II $\lambda$4686 emission line to slight changes in the density of the
secondary stellar wind in its acceleration zone.
The increase in density might result from the influence of the enhanced
X-ray emission arising from the shocked secondary wind further downstream, and/or from
accreting mass from the primary stellar wind.

A monotonic change after periastron passage might result from the accretion event.
Although not much mass is accreted during the $\sim 10~$week long accretion phase,
it is enough to change the structure of the outer envelope of the secondary star.
The secondary itself is still recovering from the 19th century Great Eruption,
when it accreted more than $10 M_\odot$ (Soker 2007b).
Therefore, it is possible that the secondary requires several years to reestablish
its structure, and before it managed to do so it encounters the next accretion phase.

Soker \& Behar (2006) speculated that the high velocity component appearing very close to
periastron passage results from accreted primary wind material by the secondary star.
At the beginning of the accretion phase the accreted gas has a relatively high
specific angular momentum, and might collimate fast segments of the secondary wind.
This might explain the new wind component mentioned by Nielsen et al. (2007)
to appear near periastron.

Despite the above description, it will be much easier to account for the velocity change
along the orbit if we observe $\eta$ Car such that the secondary is closer to
us near periastron, namely, $\gamma \sim 0$ (see Figure \ref{orbitf}).
This implies that the secondary moves away from us after periastron and toward us
before periastron.
This is opposite to the orbit orientation suggested by Nielsen et al. (2007).
In both cases the orbital plane is tilted by $\sim 45 ^\circ$ to the line of sight,
e.g., Davidson et al. (2001) give the value $i \simeq 41^\circ$, where $i$ is the angle
between a line perpendicular to the orbital plane and the line of sight.
We therefore turn to study the blueshift expected from lines emitted or absorbed in
the acceleration zone of the secondary wind.

\section{LINES VELOCITY}
\label{velocity}

\begin{figure}
\resizebox{0.89\textwidth}{!}{\includegraphics{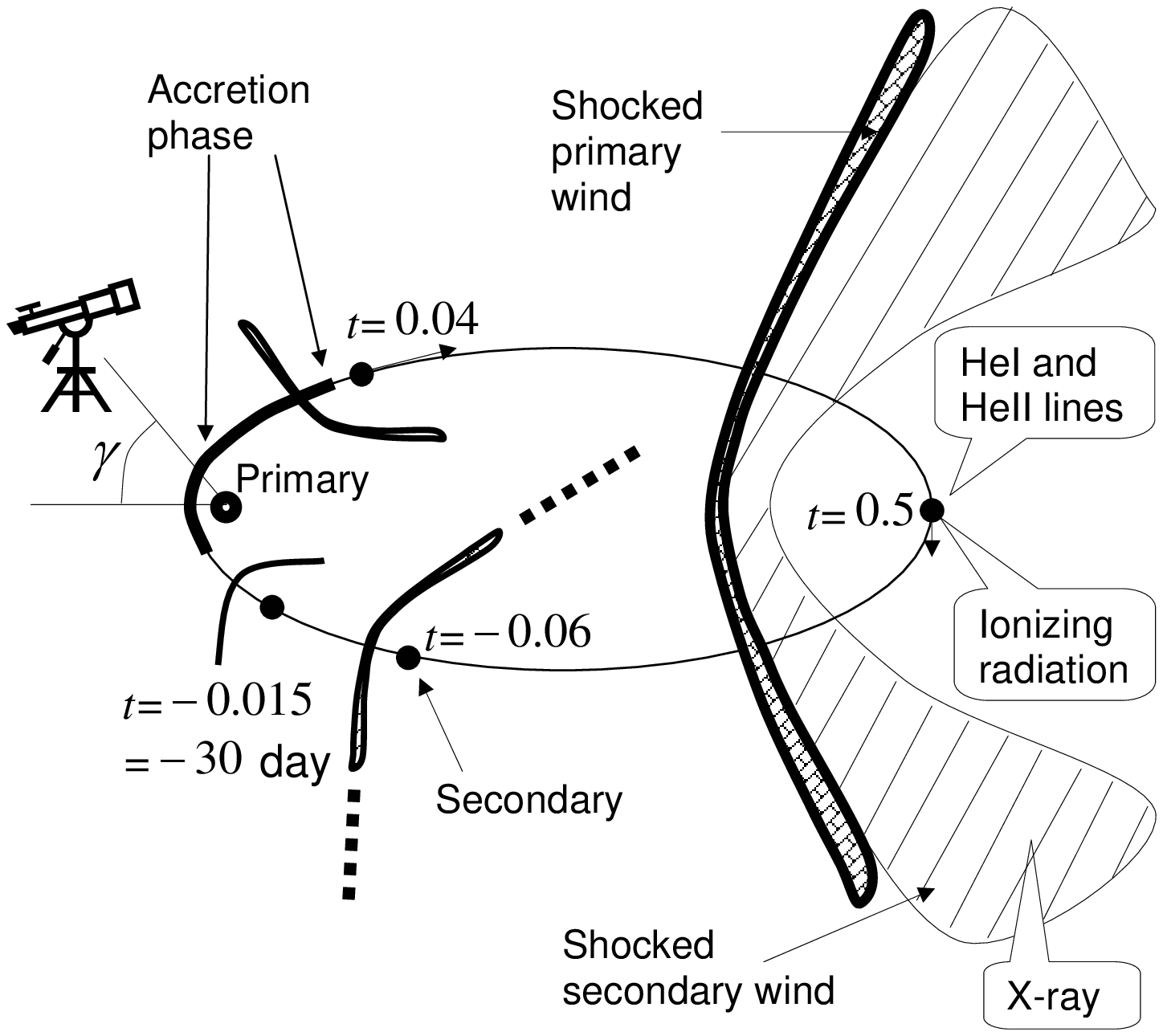}}
\vskip -7.2 cm
\caption{The orbital plane and a cut through the primary wind shocked region;
this region extends beyond the size of the figure, as drawn schematically at phase
$t=-0.06$.
The phases are marked at four points along the orbit.
The elliptical orbit, the primary stellar size, the distance of the shocked
primary wind region and its width, are drawn approximately to scale for
$e=0.9$ (but not the secondary star or the secondary wind shock region).
The thick section of the orbit is approximately where accretion phase
occurs according to the model (Soker 2005; Akashi et al. 2006).
At phase $t=0.5$ we mark the sources of the X-ray emission (the shocked secondary wind),
of the ionizing radiation (the secondary photosphere), and of the HeI and HeII lines
(the acceleration zone of the secondary stellar wind).
The orbital plane is tilted by $i \simeq 45^\circ$ to the plane of the sky. }
\label{orbitf}
\end{figure}

\begin{figure}
\resizebox{0.89\textwidth}{!}{\includegraphics{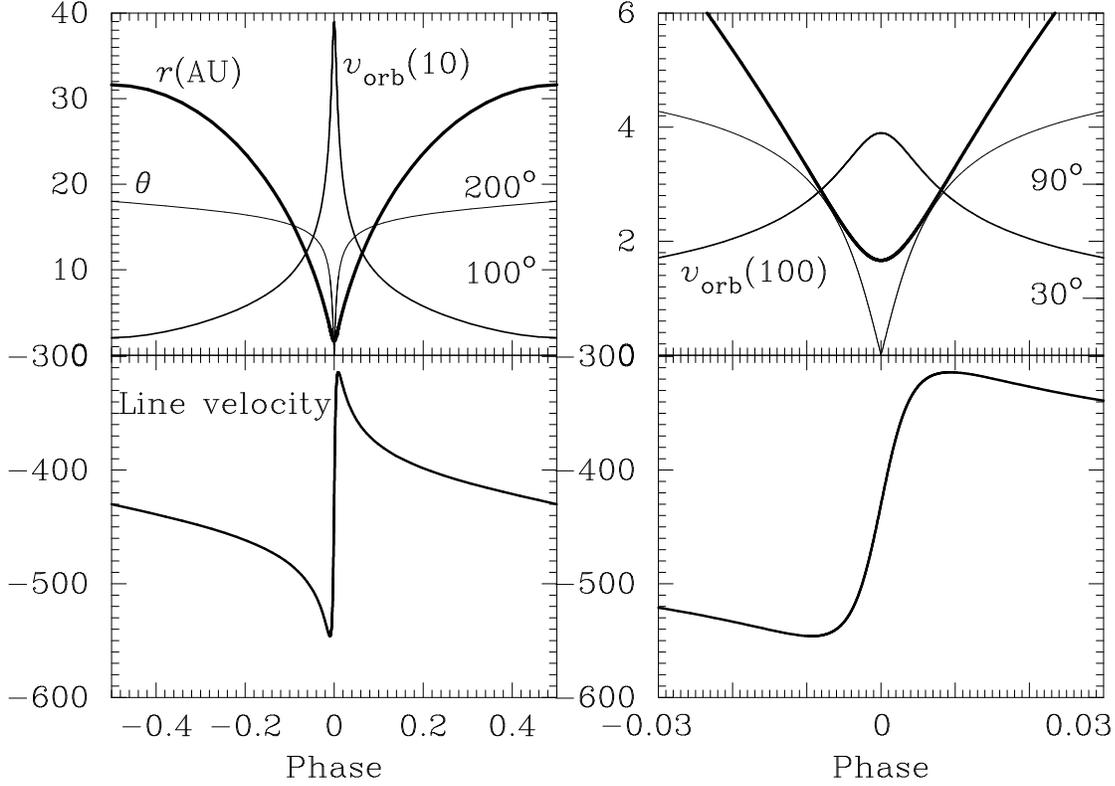}}
\caption{Several physical variables as a function of orbital phase (phase zero is at periastron),
for $e=0.9$.
The left column covers the entire orbit, while the column on the
right covers the time just prior to and after periastron passage.
Top row: Orbital separation (in AU) and relative orbital speed of the two stars
(in $10 \km \s^{-1}$ on the left and $100 \km \s^{-1}$ on the right).
The angle $\theta$ is the relative direction of the two stars as measured from periastron
(scale on the right in degrees).
Bottom row: The variation of the blueshifted line speed with phase (in $\km \s^{-1}$.
The line is formed in the acceleration zone where the secondary wind speed
relative to the secondary star is $v_{\rm zone} =430 \km \s^{-1}$, the orientation angle is
$\gamma=0$, and the inclination is $i=45 ^\circ$. }
\label{binary}
\end{figure}

The $\eta$ Car binary parameters used by us are as in the previous
papers in this series (Soker 2005; Akashi et al.\ 2006; Soker \&
Behar 2006; Soker 2007a), and are compiled by using results from
several different papers
(e.g., Ishibashi et al. 1999; Damineli et al. 2000; Corcoran et
al. 2001, 2004; Hillier et al. 2001; Pittard \& Corcoran 2002;
Smith et al. 2004; Verner et al. 2005).
The assumed stellar masses are $M_1=120 M_\odot$, $M_2=30 M_\odot$, and
orbital period 2024 days, hence the semi-major axis is $a=16.64 \AU$.
The winds properties are not directly relevant to us, but
the secondary wind mass loss rate and velocity are similar to other stars having
the HeII and HeI lines (see section 2); these are
$\dot M_2 \simeq 10^{-5} M_\odot \yr^{-1}$ and $v_2 \simeq 3000 \km \s^{-1}$,
respectively.
The binary system configuration is presented in Figure \ref{orbitf},
while in Figure \ref{binary} we present the variation of the relative speed
between the two stars $v_{\rm orb}$, of the orbital separation $r$,
and of the orbital angle $\theta$, with phase (equals zero at periastron),
for $e=0.9$.
The angle $\theta$ is the relative direction of the secondary to the primary
as measured from periastron.

The blueshifted line velocity is given by
\begin{equation}
v_{\rm line} = v_m  -  v_{\rm zone}
\label{vline1}
\end{equation}
where $v_{\rm zone}$ is the velocity of the wind's zone where the line
forms relative to the secondary star, and $v_m$ is the modulation
resulting from the orbital motion.
For a given orbital period and orientation angle $\gamma$, the blueshifted line
velocity dependance on the binary masses, inclination angle $i$, and a
large eccentricity $e$ is given by
\begin{equation}
v_m \propto \frac{M_1}{(M_1+M_2)^{2/3}} (1-e)^{-1/2} \sin i.
\label{shock1}
\end{equation}
the dependance on the masses and inclination angle is accurate, while the
dependance on eccentricity is a good approximation for large eccentricity.
In the lower row of Figure \ref{binary} we present the line of sight velocity of a
spectral line formed where $v_{\rm zone}=430 \km \s^{-1}$, for $i=45 ^\circ$, $e=0.9$, and
$\gamma=0$.

\section{FITTING THE LINE VELOCITY CURVE}
\label{orientation}

In this section we change the orientation angle $\gamma$ (see Figure 1), the
inclination angle $i$, and the eccentricity $e$ to fit the results of Nielsen et al. (2007).
In the figures below the fitted line velocity from Nielsen et al. (2007) is drawn
as a solid thick line.
We found that in all case $v_{\rm zone}= 430 \km \s^{-1}$ (see eq. \ref{vline1}) gives the
best results, and we will use this value in all cases.
In principle we could play also with the masses of the two stars, hence with the
orbital semimajor axis.
However, because the line speed goes as $\sim M_1^{1/3}$ (eq. \ref{shock1}),
for a small change in the line speed a large change in the primary mass is required,
mainly higher mass than $M_1=120 M_\odot$ is required, while most papers cite this
or lower masses (e.g., Corcoran 2007).
Lower secondary mass than we use is also problematic (Verner et al. 2005).
Therefore, we don't change the masses of the stellar components.

We fitted the data points from Nielsen et al. (2007), rather than the curve
they draw through their points.
However, for clarity, in the figures below we draw their curve instead of their data.
For that, the fit does not appear perfect.
In addition, we will not give high weight to an exact fitting very close to periastron passage.
The reason is that very close to periastron the properties in the acceleration zone of the
secondary wind can change because of the X-ray emission and accretion (Soker \& Behar 2006).
Soker \& Behar (2006) argued that such property variations can cause the
He~II $\lambda$4686 line to be formed in a region with a higher velocity near periastron.
For that, in all graphs below there is no perfect fitting very close to phase zero.

\begin{figure}
\resizebox{0.49\textwidth}{!}{\includegraphics{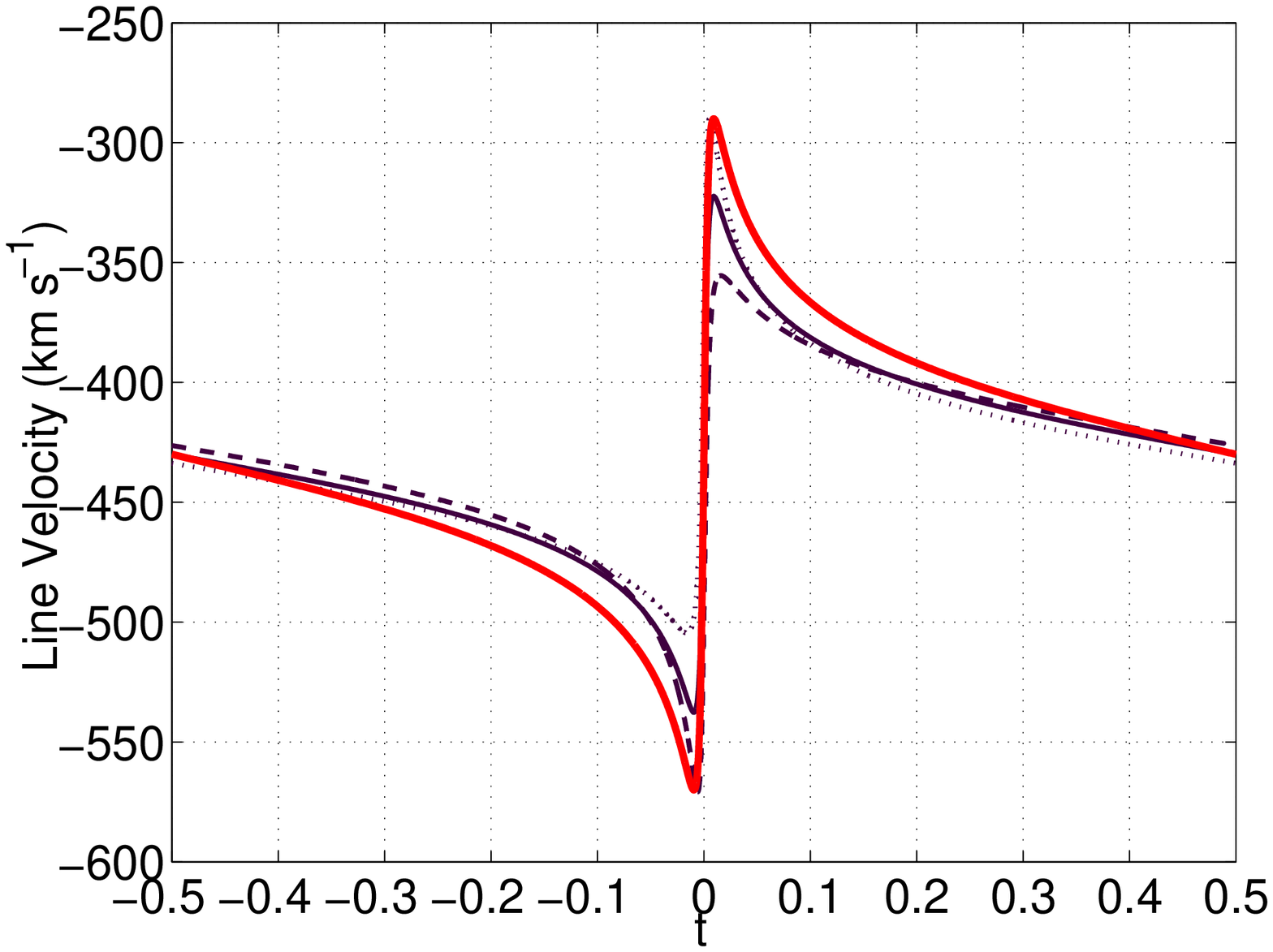}}
\resizebox{0.49\textwidth}{!}{\includegraphics{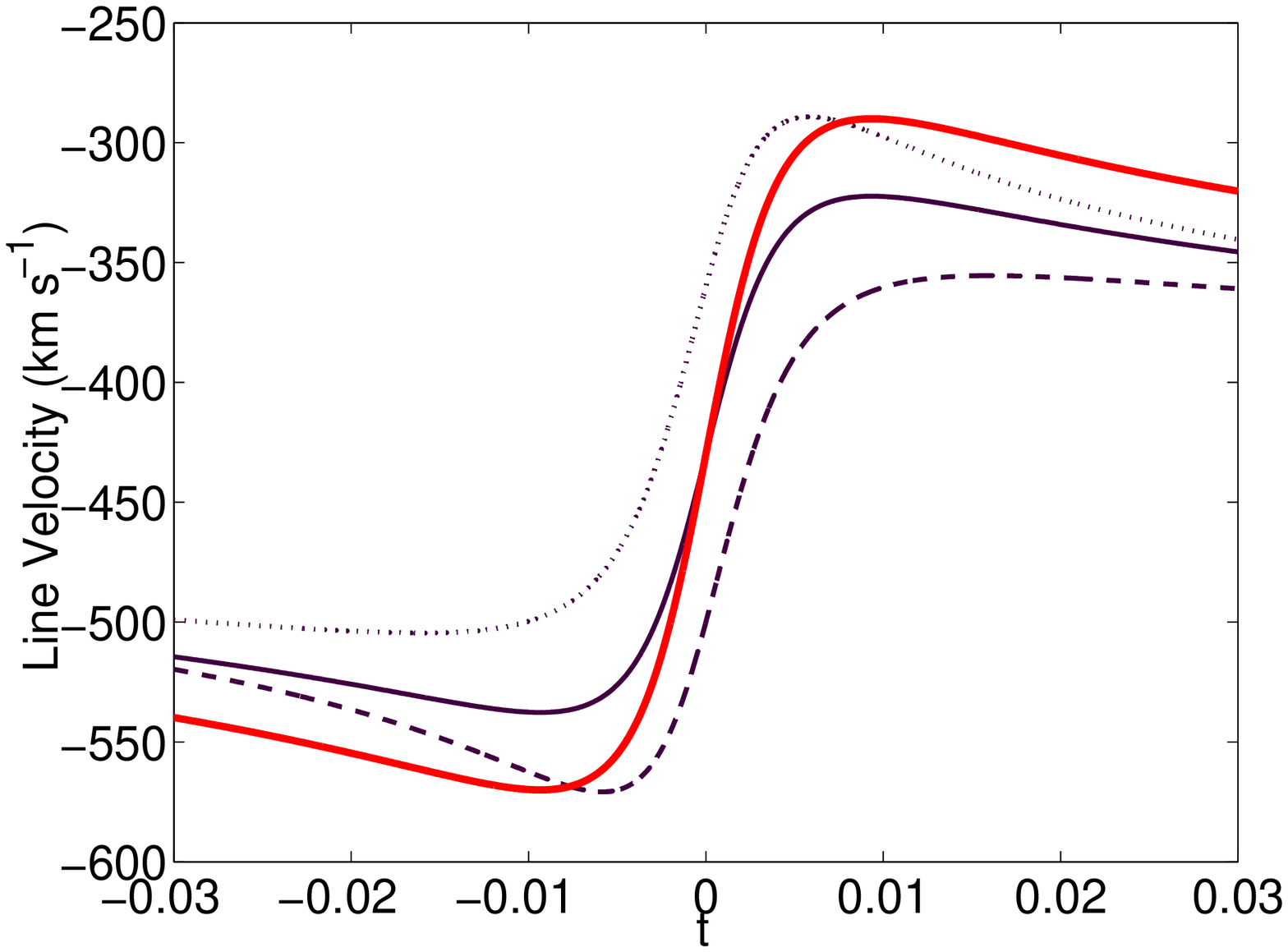}}
\caption{
The HeI lines velocity fitted by Nielsen et al. (2007; solid thick line)
is compared with the model discussed here for an inclination angle
(between the orbital axes, which is perpendicular to the orbital plane,
and line of sight)
of $i=41^\circ$ and an eccentricity of $e=0.9$.
Calculations for three orientation angles (see Figure 1) are presented with thin lines:
$\gamma=-20^\circ$ (dotted); $\gamma=0$ (solid); $\gamma=20^\circ$ (dashed).
The left side covers the entire orbit, while that on the
right covers the time just prior to and after periastron passage. }
\label{vline4f}
\end{figure}

We start by taking $i = 41^\circ$ (Davidson et al. 2001), $e = 0.9$ as was used in previous
papers in this series, and $\gamma = 0$.
As evident from the thin solid line in Figure \ref{vline4f}, these parameters give line
velocity with amplitude lower by $\sim 30 \km \s^{-1}$ than the line velocity
found of Nielsen et al. (2007).
When comparing the observations (Nielsen et al. 2007) over the entire period,
we find that for these parameters our modelled line velocity is lower
by $\sim 20 \km \s^{-1}$ than required.
By varying the orientation of the semimajor axis to $\gamma \simeq 10^\circ$ or $\gamma \simeq -10^\circ$
we can make a better fitting either before or after the spectroscopic event, respectively,
but not in both epochs.

\begin{figure}
\resizebox{0.49\textwidth}{!}{\includegraphics{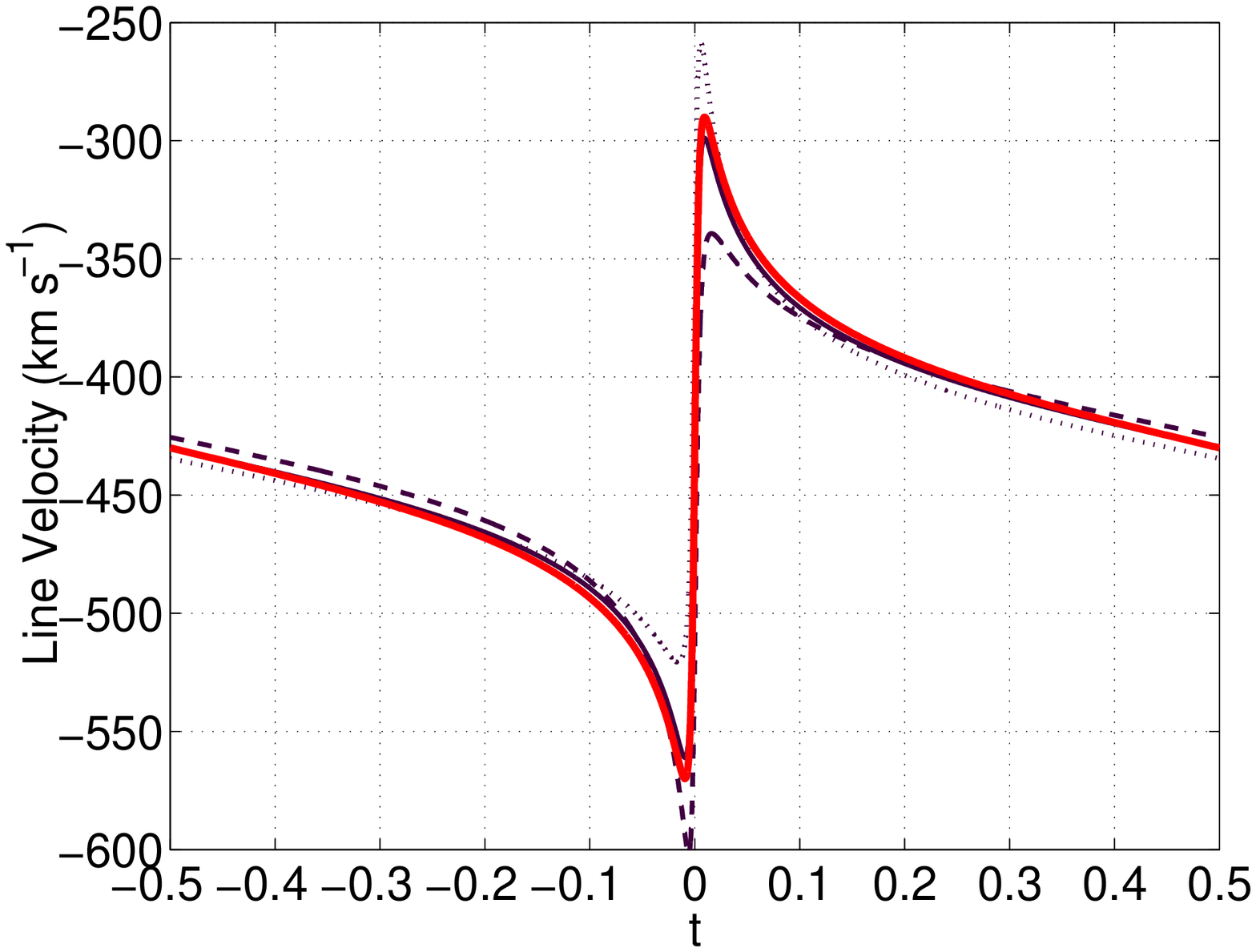}}
\resizebox{0.49\textwidth}{!}{\includegraphics{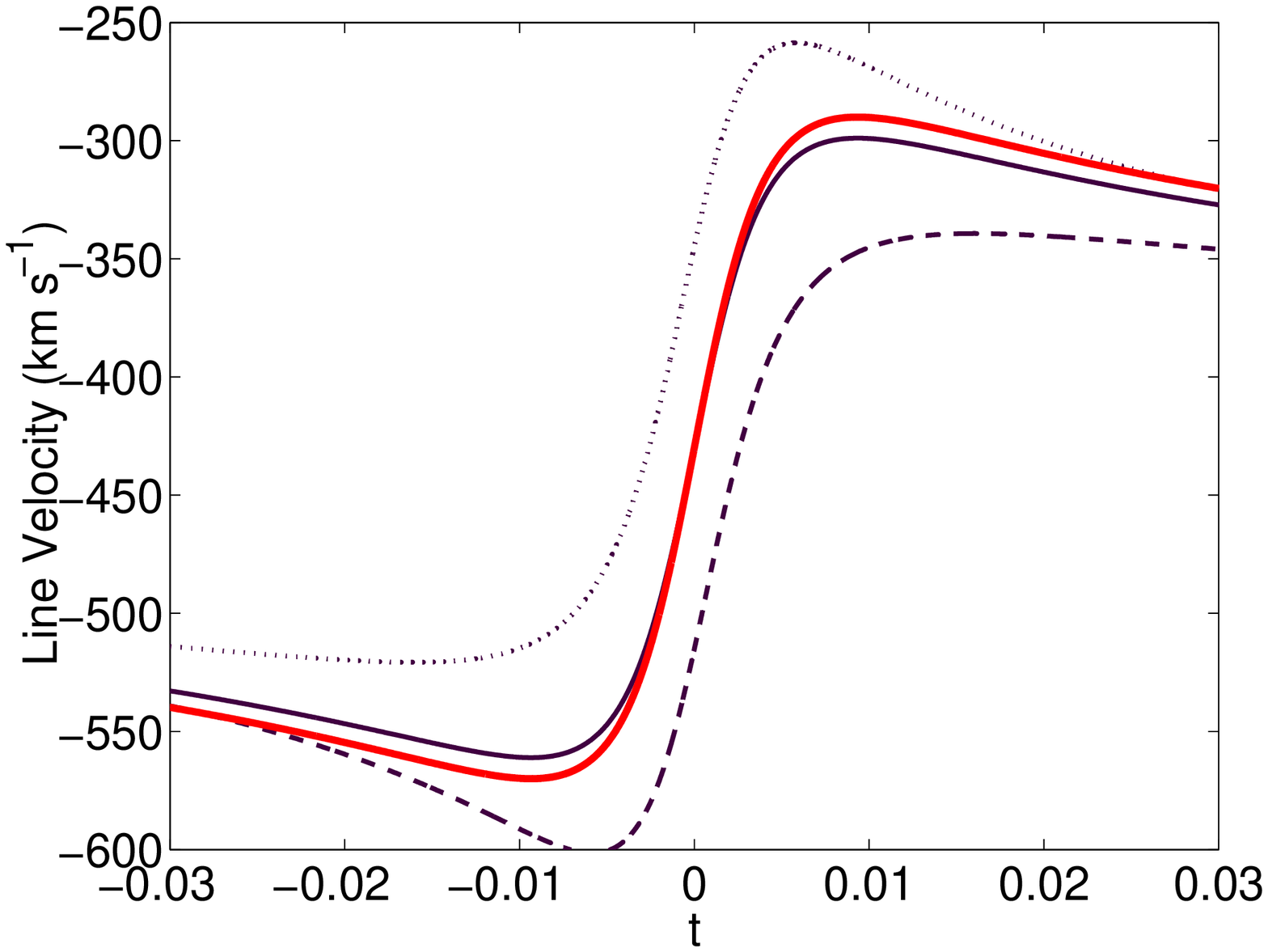}}
\caption{
Like Fig. \ref{vline4f}, but for $i=53^\circ$ and $e=0.9$. }
\label{vline1f}
\end{figure}

Next we try to increase the inclination angle $i$, while keeping the other parameters unchanged.
The dependance of the velocity amplitude on $i$ is according to equation (\ref{shock1}).
We find the best fit to be for $ i \simeq 53^\circ$.
This is somewhat larger than the value considered in some previous papers
(e.g. Davidson et al. 2001), but not in others (e.g., Corcoran 2007).
The line velocity for this case is drawn in Figure \ref{vline1f}.

\begin{figure}
\resizebox{0.49\textwidth}{!}{\includegraphics{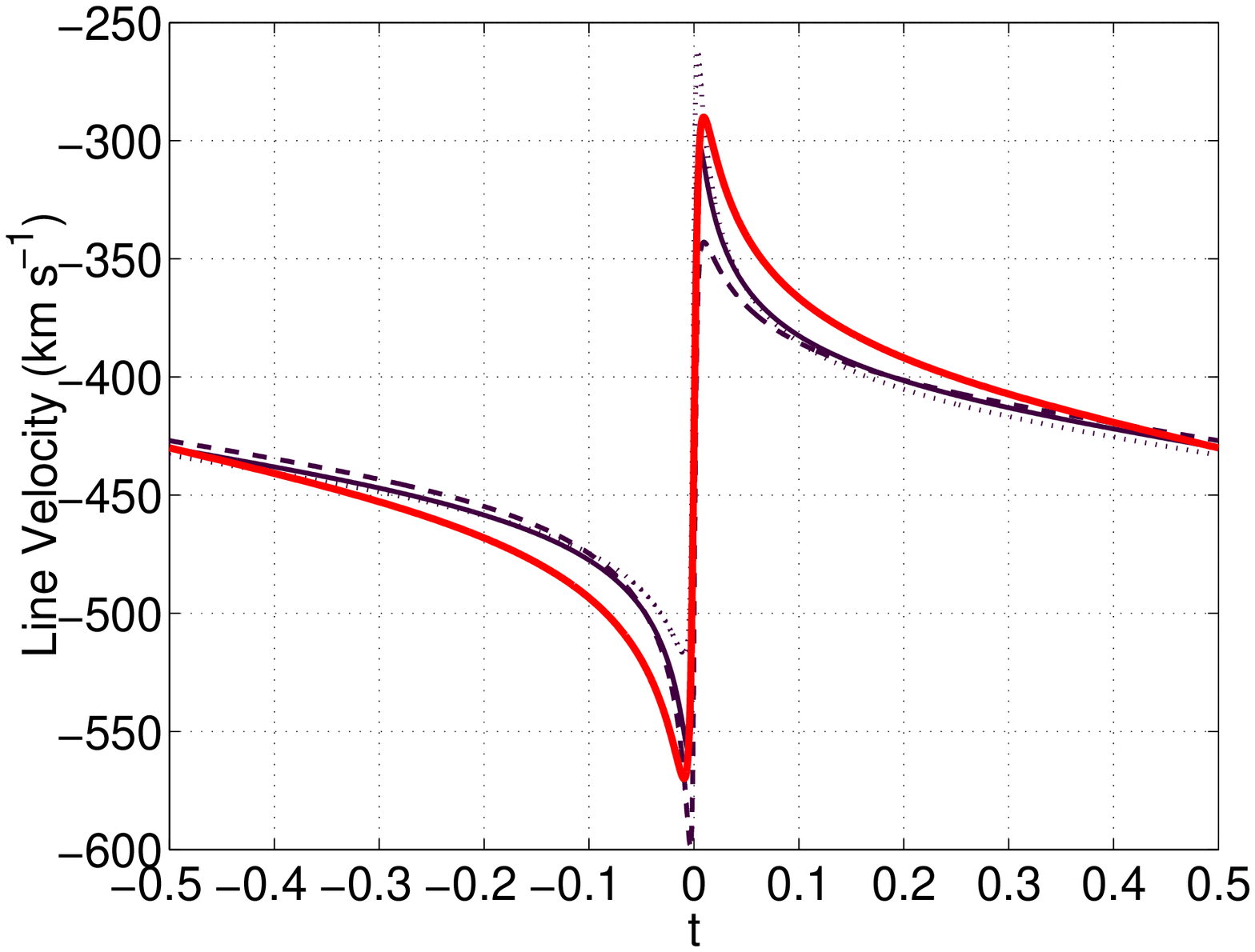}}
\resizebox{0.49\textwidth}{!}{\includegraphics{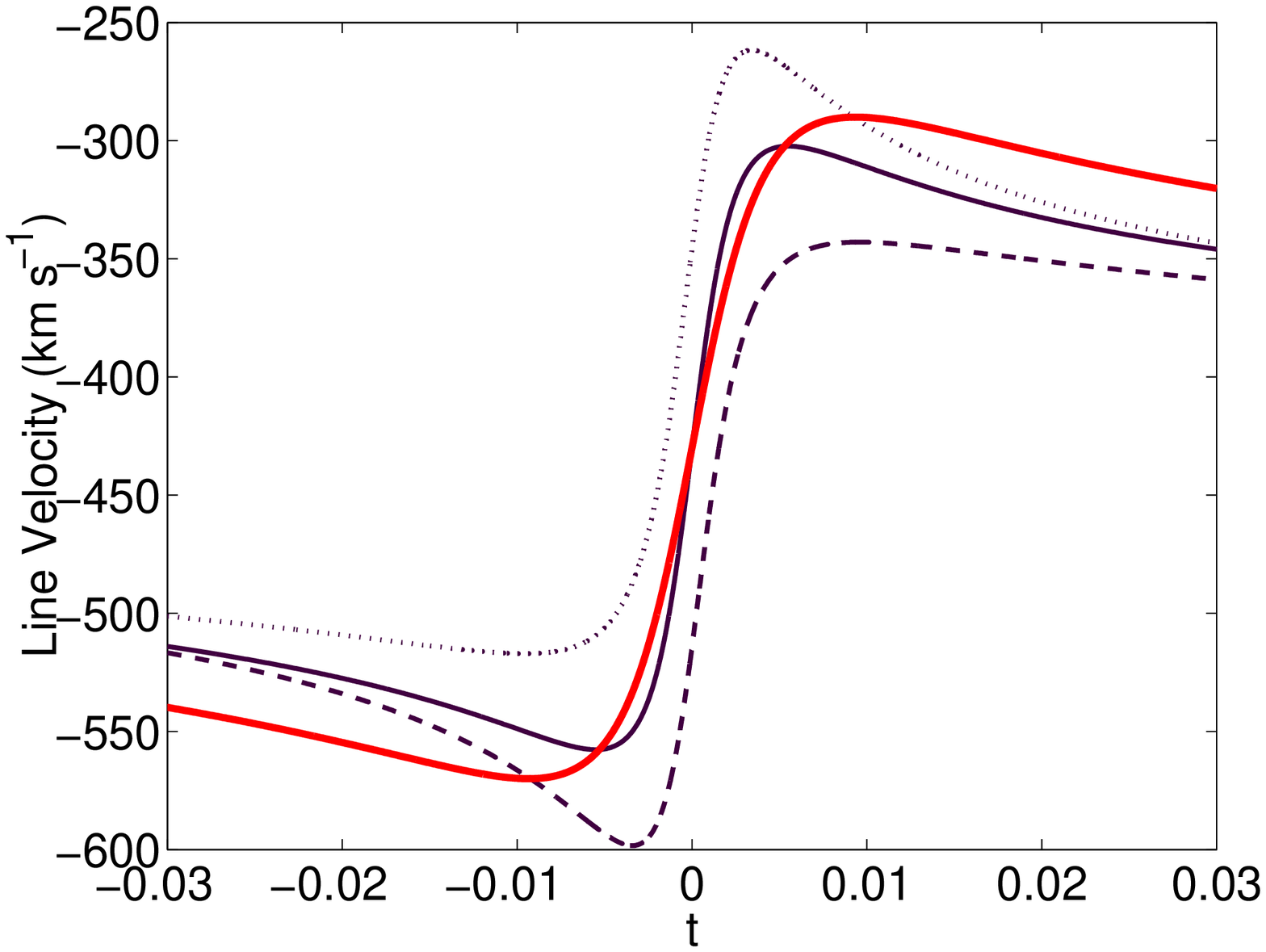}}
\caption{
Like Fig. \ref{vline4f}, but for $i=41^\circ$ and $e=0.93$ }
\label{vline2f}
\end{figure}

\begin{figure}
\resizebox{0.49\textwidth}{!}{\includegraphics{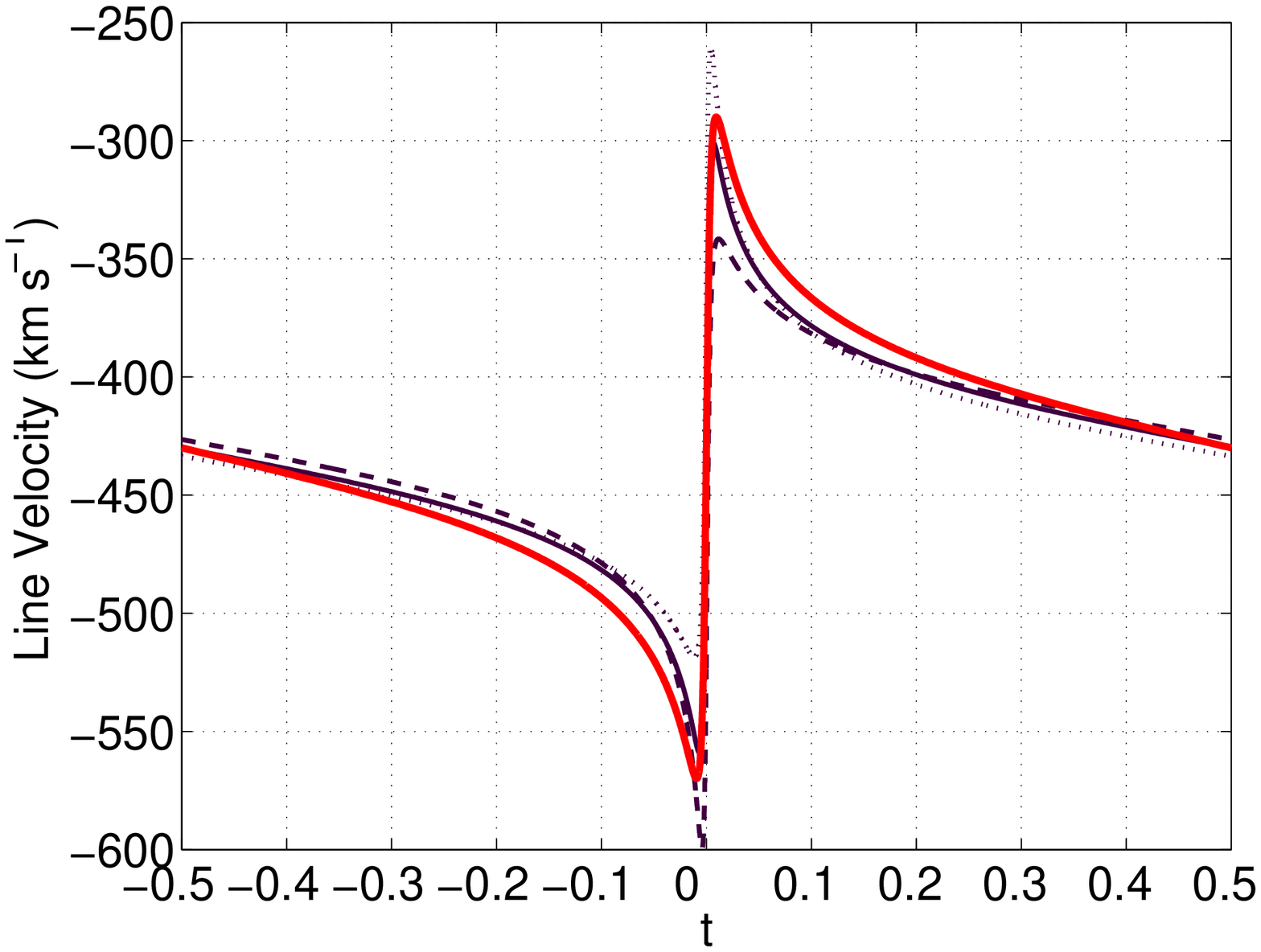}}
\resizebox{0.49\textwidth}{!}{\includegraphics{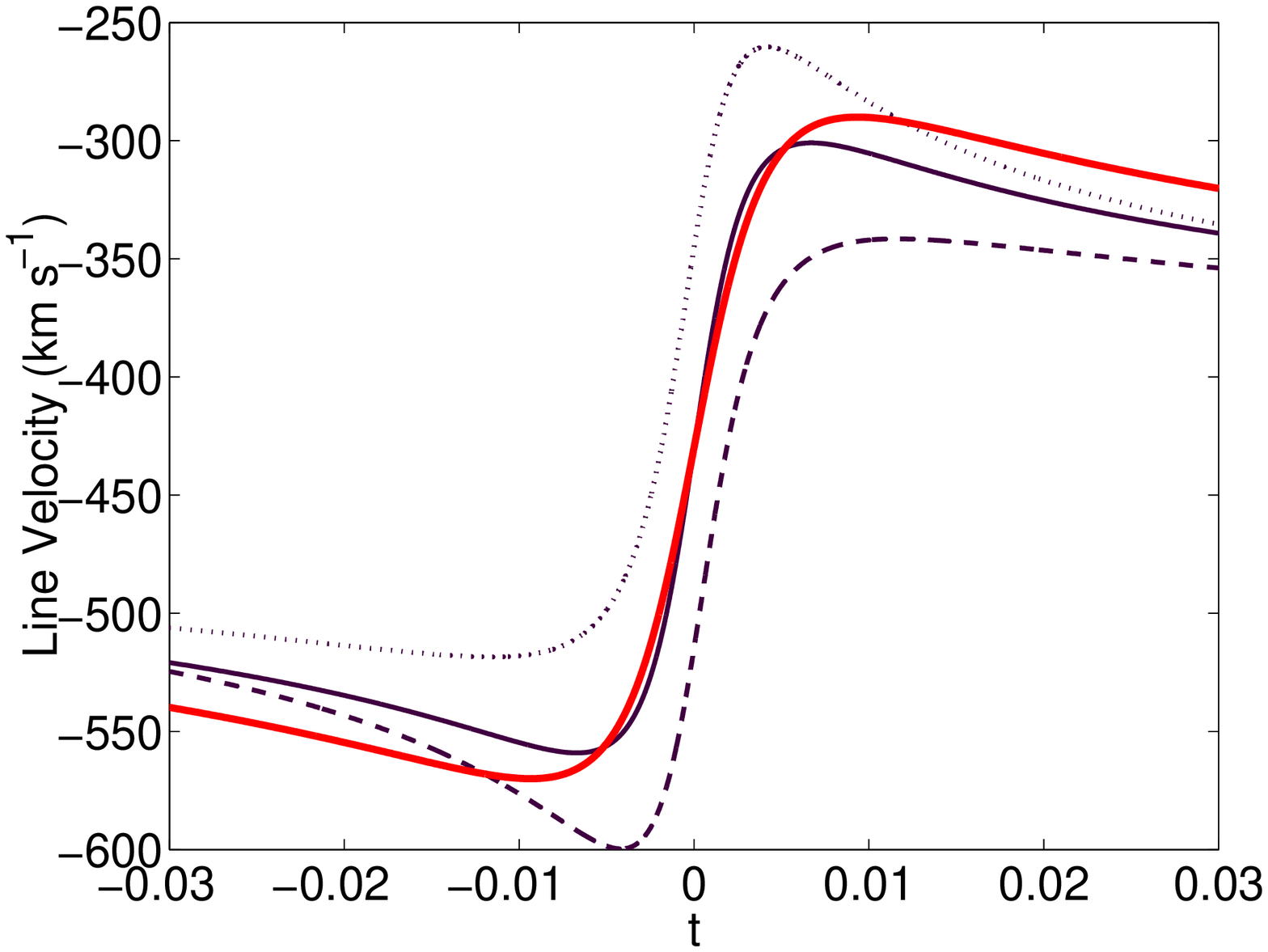}}
\caption{
Like Fig. \ref{vline4f}, but for $i=45^\circ$ and $e=0.92$ }
\label{vline3f}
\end{figure}

Another option to get a better fit is to increase the eccentricity $e$.
Keeping all parameters unchanged, namely, $i = 41^\circ$ etc.,
we find the best fitting for $e \simeq 0.93$, as expected from eq 2.
The result for this case is plotted in Figure \ref{vline2f}.
We also try an intermediate case, with values of $e=0.92$ and $i=45^\circ$.
The result for this case is plotted in Figure \ref{vline3f}.

In all cases we find that even a small variation as $20^\circ$ from $\gamma=0$
decreases the fitting.
We conclude that the semimajor axis orientation angle is $-10 ^\circ \la \gamma \la 10^\circ$.


\section{DISCUSSION AND SUMMARY}
\label{summary}

We postulated that the HeI lines are formed in the acceleration zone of the
secondary wind, rather than in the primary wind as suggested by
Nielsen et al. (2007).
Supporting arguments for this assumption are given in Section 2.
Under this postulate, the Doppler shift of the HeI lines and its variation with the
orbital phase is caused by the orbital motion of the secondary around the center of
mass of the binary system.
We found that using binary parameters from the literature we can nicely fit the
Doppler shift.
In section 3 we described the binary system and the calculation of the Doppler shift.
In section 4 we fitted the calculated Doppler shift to the lines velocity
observed by Nielsen et al. (2007) by varying the inclination angle $i$, the eccentricity $e$,
and the orientation angle of the semimajor axis $\gamma$.
In all cases we found that the best fit is obtained if the HeI lines are formed in the
acceleration zone of the secondary wind where the wind speed is
$v_{\rm zone}= 430 \km \s^{-1}$ (see eq. \ref {vline1}).

In all cases the best fit is for an orientation angle $\gamma =0$.
Namely, the secondary is closer to us at periastron passage with
$-10^\circ \la \gamma \la 10^\circ$.
Falceta-Goncalves et al. (2005) already argued for this orientation.
However, they based their claim on a model for the X-ray light curve which we
disagree with (Akashi et al. 2006).
Others (e.g., Nielsen et al. 2007; Corcoran 2007) prefer the orientation to
be opposite to our finding, namely, $\gamma=180^\circ$.
Smith et al. (2004) used $e=0.8$, and argued for a perpendicular orientation
with $\gamma= 90 ^\circ$.
The orientation angle is one of the binary parameters in high priority
for the community to agree upon.

According to our results, and considering the several sources of uncertainties
(both in model and observations of the lines velocities), the eccentricity is in
the range $0.90 \la e \la 0.95$,
and the inclination angle in the range $40^\circ \la i \la 55 ^\circ$.
Lower values of $e$ require higher values of $i$.
Because the inclination is constraint by other observations (e.g., Davidson et la. 2001)
to $i \simeq 41-45^\circ$, we prefer the range $0.92 \la e \la 0.95$
This eccentricity is higher than the value of $e=0.9$ used in the papers dealing with the
accretion model near periastron (Soker 2005; Akashi et al. 2006), and implies closer
approach of the two stars by $\ga 20\%$. This substantially increase the likelihood of
the accretion model.

{{{ We have the following answers to comments raised by some colleagues
against our assumption that the He~I lines originate in the secondary wind.
(1) Some hydrogen lines with similar profiles (Nielsen et al. 2007) originate in the primary wind.
Answer: We note that the P-Cygni hydrogen line profile, e.g., H I $\lambda$4103, does not
change as the He I lines do (Nielsen et al. 2007).
(2) The coincidence that the secondary wind velocity where the
lines are formed,
 $v_{\rm zone} = -430 \km\s^{-1}$, is practically the same as the primary wind
  terminal velocity.
Answer:  Although this is a coincidence, the same can be said on the model where the He I
  lines originate in the primary wind. How come the region where the
  lines are formed in the primary wind changes its velocity exactly as the
  secondary speed around the center of mass does?
(3) Other emission lines expected from the secondary are not observed, e.g.,
the He~II $\lambda$4686 line.
Answer: the secondary wind must be highly clumpy,
with dense and somewhat cooler blobs emitting the He I lines.
The sensitivity of lines intensities to the exact wind properties can account
for the non-presence of other expected lines.
(4) $\eta$ Car B is much fainter than $\eta$ Car A, and it would be difficult
to see the emission from B.
Answer: The secondary is hot and its luminosity is $\sim 20\%$ that of the primary.
This seems enough to account for the He I lines from the secondary to be detected.
We note that in the model of Nielsen et al. (2007), the secondary will ionize
a small section of the primary wind, and it is hard to
explain the strong emission.
(5) The secondary radiation will ionize the primary wind, and we expect He I lines
 from the primary wind.
Answer: We agree with that. However, we expect the contribution from the primary wind to be small,
 as stated in point (4) above.

What we have shown here is that {\it if} the lines originate in the acceleration zone
of the secondary wind, then their Doppler shift can be easily explained. This is not
the case if the lines originate in the primary wind.
In any case, we stress again that He I lines are expected from $\eta$ Car B.
Fullerton et al. (1996) listed several stars with effective temperature of
$\sim 36,000-40,000 \K$ that have the He~I 5876 line in absorption.
The line velocity is typically much lower than the wind terminal speed.
These lines are formed in the wind; for example, Rauw et al. (2001) mentioned
that the HeI lines are formed in the wind of HD 192639, which has an effective
temperature similar to that of $\eta$ Car B (Fullerton et al. 1996).

Another possibility is that the He~I lines originate near the stagnation point,
namely, in the post shock primary wind.
The stagnation point speed around the center of mass is $\sim 70 \%$ that of
$\eta$ Car B.
This possibility requires a higher eccentricity, $e \simeq 0.95$ for $i = 53^\circ$,
or a higher inclination, $i \simeq 70^\circ$ for $e = 0.93$, or some combined increased
in their values.
 }}}

If the binary parameters we found here are correct, then they have implications for
Doppler shift in other wavelengths as well.
Consider X-ray emission. Most of the hard X-ray emission comes from the
shocked secondary wind near the stagnation point of the colliding winds
(Pittard et al. 1998; Corcoran et al. 2001; Pittard \& Corcoran 2002;
Akashi et al. 2006; Hamaguchi et al. 2007).
This region is close to the secondary star and moves with it.
Therefore, a Doppler shift is expected in the X-ray band.
Indeed, very near periastron passage, phase $ < 0.03$, the peak in the X-ray spectrum
is blueshifted by $\sim 150-250 \km \s^{-1}$ (Behar et al. 2006).
No blue shift is observed about a year before periastron.
We therefore suggest that this shift in the peak is due mainly to the orbital
motion of the secondary. Some contribution can come from a second wind component
appearing near periastron (Behar et al. 2006).
Corcoran (2007) took $\gamma=180^\circ$ and proposed that the changes in the
Doppler shift of the X-ray lines with orbital phase result from the change in
the orientation of the colliding winds region.
In principle such a model can work when we observe the system in the orbital plane,
namely $i=90^\circ$. However, if we take the inclination into account, then even
during apastron passage a blueshift component of $v_a \simeq \sim v_p \cos i$ is expected,
where $v_p$ is the (blueshifted) velocity causing the blueshift the model supposes to explain near periastron.
Taking $i \la 55^\circ$ for $\eta$ Car, we find that $v_a \ga 0.5 v_p$.
Such a blue shift is not observed near periastron (Behar et al. 2006; Corcoran 2007).

{{{ We thank John Hillier, Ted Gull, and Michael Corcoran for pointing out
the difficulties with the assumption that the He I lines originate in the secondary wind. }}}
This research was supported by a grant from the
Asher Space Research Institute at the Technion.

\end{document}